\documentstyle[12pt,amssymb]{article}
\begin{document}
\def\be{\begin{equation}}
\def\ee{\end{equation}}
\def\H{\Bbb H}
\def\K{\Bbb K}
\def\R{\Bbb R}
\def\C{\Bbb C}
\def\Z{\Bbb Z}
\def\d{\partial}

\renewcommand{\theequation}{\thesection{.\arabic{equation}}}
\rightline{\today}
\rightline{CERN-TH/98-141}
\rightline{LAPTH-681/98}
%\rightline{hep-th/9805149}

\vfill

\centerline{\Large \bf
Explicit solution of the quantum} 
\medskip 
\centerline{\Large \bf 
three-body Calogero-Sutherland model}
\vskip 1cm
\centerline{\large {\sc A.M. Perelomov}$^{a,}$\footnote{On leave of
absence 
from Institute for Theoretical and Experimental 
Physics, 117259 Moscow, Russia. Current e-mail address: 
perelomo@posta.unizar.es.}, {\sc E. Ragoucy}$^{b,}$\footnote{On leave 
of absence from
    Laboratoire de Physique Th{\'e}orique LAPTH, 74941 Annecy-le-Vieux,
    France. E-mail address: ragoucy@lapp.in2p3.fr}, and {\sc Ph.
Zaugg}$^a$}
\medskip\noindent
$^a${\sl Laboratoire de Physique Th{\'e}orique LAPTH,
Chemin de Bellevue, BP 110, 74941 Annecy-le-Vieux Cedex, France}\\
\medskip\noindent
$^b${\sl Theory Division, CERN, 1211 Geneva 23, Switzerland}

\vfill

\centerline{\bf Abstract}

{\small\noindent
  The class of quantum integrable systems associated with root systems
  was introduced in \cite{OP77} as a generalization of the
  Calogero--Sutherland systems \cite{Ca,Su}.  In a recent 
  note \cite{Pe98}, it was proved that for such systems with a potential 
  $v(q)=\kappa (\kappa -1)\,\sin ^{-2}q$, the series in the product of 
  two wave functions is the $\kappa $-deformation of the Clebsch--Gordan 
  series. 
  This yields recursion relations for the wave functions of those systems 
  and, related to them, for generalized zonal spherical functions 
  on symmetric spaces. \\
  In the present note, this approach is used to compute the explicit 
  expressions for the three-body Calogero--Sutherland wave functions,
which 
  are the Jack polynomials.  We conjecture 
  that similar results are also valid for the more general
  two-parameters deformation ($(q,t)$-deformation) introduced by 
  Macdonald \cite{Ma}.}

\vfill

\leftline{CERN-TH/98-141}
\leftline{May 98}
\newpage

\section{Introduction}

The class of quantum integrable systems associated with root systems
was introduced in \cite{OP77} (see also \cite{OP78}) as generalization
of the Calogero--Sutherland systems \cite{Ca,Su}.  Such
systems depend on one real parameter $\kappa$ (for the type $A_n$,
$D_n$ and $E_6, E_7, E_8$),
on two parameters (for the type $B_n, C_n, F_4$ and $G_2$) and on
three parameters for the type $BC_n$. These parameters are related to
the coupling constants of the quantum system.

For special values of this parameter $\kappa$, the wave functions
correspond to the characters of the compact simple Lie groups ($\kappa
=1$) \cite{We}, or to zonal spherical functions on symmetric
spaces ($\kappa =1/2, 2, 4$) \cite{Ha,Hel}.  At arbitrary values of
$\kappa$, they provide an interpolation between these objects.

This class has many remarkable properties. Let us only mention that
the wave functions of such systems are a natural generalization of
special functions (hypergeometric functions) to the case of several
variables. The history of the problem and some results can be found in
\cite{OP83}. It was recently shown in \cite{Pe98}, that the product of
two wave functions is a finite linear combination of analogous
functions, namely of functions that appear in the corresponding
Clebsch--Gordan series.  In other words, this deformation
($\kappa$-deformation) does not change the Clebsch--Gordan series. For
rank 1, one obtains the well-known cases of the Legendre, Gegenbauer
and Jacobi polynomials and the limiting cases of the Laguerre and
Hermite polynomials (see for example \cite{Vi}). Some other cases were
also considered in \cite{Her,Jac,Ko,Jam,Se,Vr76,Vr84,Ma,La} and
\cite{St}.
In this note we use this property in order to obtain the explicit
expressions for the Jack polynomials\footnote{We use the name of Jack
polynomials, although, strictly speaking, they are slightly
different from those introduced by Jack \cite{Jac}. Another possible
denomination is generalized Gegenbauer polynomials \cite{Jam}.}
of type $A_2$ 
which give the solution of the three-body Calogero--Sutherland model. 
For special values of $\kappa =1/2, 2, 4$ we obtain the explicit 
expressions for zonal polynomials of type $A_2$.

We conjecture that these results remain valid for the Macdonald 
polynomials of type $A_2$ \cite{Ma} and this will be the subject of a 
separate communication \cite{PRZ}.

\section{General description}
\setcounter{equation}{0}

The systems under consideration are described by the Hamiltonian (for more 
details see \cite{OP83}):
\be
H=\frac{1}{2}\,p^2 + U(q),\qquad p^2=(p,p)=\sum _{j=1}^l\,p_j^2,
\ee
where $p=(p_1,...,p_l),\,\,p_j=-i\,\frac{\partial }{\partial q_j}$, 
is a momentum vector operator, and $q=(q_1,...,q_l)$ is a coordinate
vector 
in the $l$-dimensional vector space $V\sim {\Bbb R}^l$ with standard 
scalar product $(\alpha ,q)$. They are a generalization of 
the Calogero--Sutherland systems \cite{Ca,Su} 
for which $\{ \alpha \}=\{e_i-e_j\},\,\{e_j\}$ being a standard 
basis in $V$. The potential $U(q)$ is constructed by means of 
a certain system of vectors $R^+=\{\alpha\}$ in $V$ 
(the so-called root system): 
\be
U=\sum \limits_{\alpha \in R^+}g_\alpha ^2\,v(q_\alpha ),\qquad 
q_\alpha = (\alpha ,q), \qquad
g_\alpha ^2=\kappa _\alpha (\kappa _\alpha -1).
\ee
The constants satisfy the condition $g_\alpha =g_\beta ,\,\,
\mbox{\rm if}\,\,(\alpha ,\alpha )=(\beta, \beta )$. Such systems are 
completely integrable for five types of potential \cite{OP83}. In this
note we
consider only the $A_2$ case with potential $v(q)=\sin ^{-2}q$.

\section{The Clebsch--Gordan series}
\setcounter{equation}{0}

In this section, we recall the main results of \cite{Pe98} and
specialize them to the $A_2$ case.
The Schr{\"o}dinger equation for this quantum system with $v(q)=\sin
^{-2}q$ has the form
\be
H\,\Psi ^\kappa = E(\kappa)\,\Psi ^\kappa ;\qquad
H=-\,\Delta _2+U(q_1,q_2,q_3), \qquad
\Delta _2=\sum \limits_{j=1}^{3}\,\frac{\partial ^2}
{\partial q_j^2}
\label{schrod}
\ee
with
\be
U(q_1,q_2,q_3)= \kappa(\kappa-1)
\left( \sin^{-2}(q_1-q_2)+\sin^{-2}(q_2-q_3)+
\sin^{-2}(q_3-q_1) \right).
\ee
The ground state wave function and its energy are
\be
\Psi _0^\kappa (q)=\left( \prod _{j<k}^3 \,\sin (q_j-q_k)
\right)^\kappa, \qquad E_0(\kappa )= 8\kappa^2.
\ee
Substituting $\Psi _\lambda ^\kappa =\Phi _\lambda ^\kappa \,\Psi
_0^\kappa$ 
in (\ref{schrod}) we obtain
\be
-\Delta ^\kappa \,\Phi _\lambda ^\kappa =\varepsilon _\lambda (\kappa )\,
\Phi _\lambda ^\kappa ,\qquad
\Delta ^\kappa=\Delta _2+\Delta _1^\kappa ,\qquad
\varepsilon _\lambda (\kappa )=E_\lambda (\kappa )-E_0(\kappa ).
\ee
Here the operator $\Delta _1^\kappa $ takes the form
\be
\Delta _1^\kappa =\kappa\sum _{j<k}^3\, \cot (q_j-q_k)
\left( \frac{\partial}{\partial q_j} - \frac{\partial}{\partial
q_k}\right) .
\ee
It is easy to see that $\Delta ^\kappa $ leaves invariant the set of
symmetric polynomials in variables $\exp(2iq_j)$.
Such polynomials $m_\lambda$ are labelled by an $SU(3)$ highest
weight $\lambda=m \lambda_1+n \lambda_2$, with $m,n$ non-negative
integers and $\lambda_{1,2}$ the two fundamental weights.
In general,
\be
\Phi_\lambda^\kappa = \sum_{P^+\ni \mu \leq \lambda}
C^\mu_\lambda(\kappa) \, m_\mu, \qquad m_\mu = \sum_{\lambda' \in W
  \cdot \mu} e^{2i(q,\lambda')},
\ee
where $P^+$ denotes the cone of dominant weights, $W$ the Weyl group,
and $C^\mu_\lambda(\kappa)$ are some constants, taking care of the
wave function normalization.

The most remarkable result of
\cite{Pe98} is that the product of two wave functions is a finite sum
of wave functions (a sort of the $\kappa$-deformed Clebsch--Gordan
series) 
\be
\Phi_\mu^\kappa \,\Phi_\lambda^\kappa = \sum_{\nu \in
D_\mu(\lambda )} C_{\mu \lambda }^\nu (\kappa )
\,\Phi _\nu ^\kappa.
\label{main}
\ee
In this equation, $D_\mu(\lambda)=(D_\mu+\lambda)\cap
P^+$, where
$D_\mu$ is the weight diagram of the representation with highest
weight $\mu$. 

Since $\Phi_\mu^\kappa$ are symmetric functions of $\exp(2iq_j)$,
it is convenient to work with a new set of variables
\be
\begin{array}{l}
z_1 = e^{2i q_1}+e^{2i q_2}+e^{2i q_3}, \qquad 
z_2= e^{2i(q_1+q_2)}+e^{2i (q_2+q_3)}+e^{2i (q_3+q_1)}, \\
z_3= e^{2i(q_1+q_2+q_3)}.
\end{array}
\ee
As we are in the centre-of-mass frame ($\sum_i p_i=0$), the wave 
functions depend on two variables only, which we choose to be $z_1$
and $z_2$ (it is consistent to set $z_3=1$).
In these variables, and up to a normalisation factor, $\Delta^\kappa$
reads ($\partial_i= \partial/\partial z_i$):
\be
\Delta^\kappa =(z_1^2-3z_2)\,\partial _1^2+(z_2^2-3z_1)\,
\partial _2^2+(z_1z_2-9)\,\partial _1\partial _2+(3\kappa +1)\,
(z_1\partial _1+z_2\partial _2).
\ee
Its eigenvalues are
\be
\varepsilon_{m,n}(\kappa) = m^2+n^2+m n+3\kappa(m+n).
\ee

For the rest of this note, we will use a different normalization for
the polynomials $\Phi_\lambda^\kappa$
and denote them by $P^\kappa_{m,n}$. In \cite{Pe98}
their general structure was outlined
\be
P^\kappa_{m,n}(z_1,z_2) = \sum_{p,q} C^{p,q}_{m,n}(\kappa)\,z_1^p\,z_2^q
= z_1^m\,z_2^n + {\rm lower~terms},
\label{P-struct}
\ee
with $p+q \leq m+n$ and $p-q \equiv m-n ({\rm mod~}3)$.  The first
polynomials are easy to find:
\be
P^\kappa_{0,0}=1, \qquad
P^\kappa_{1,0}=z_1, \qquad
P^\kappa_{0,1}=z_2.
\ee
In \cite{Pe98} simple instances of (\ref{main}) for
$P^\kappa_\lambda=
P^\kappa_{1,0}$ or $P^\kappa_{0,1}$ were given
\begin{eqnarray}
z_1\,P_{m,n}^\kappa &=& P_{m+1,n}^\kappa +a_{m,n}(\kappa)\,
P_{m,n-1}^\kappa +c_{m}(\kappa) \,P_{m-1,n+1}^\kappa ,
\label{rec-z1} \\
z_2\,P_{m,n}^\kappa &=& P_{m,n+1}^\kappa +\tilde{a}_{m,n}(\kappa)\,
P_{m-1,n}^\kappa +c_{n}(\kappa) \,P_{m+1,n-1}^\kappa ,
\label{rec-z2}
\end{eqnarray}
where
\begin{eqnarray}
a_{m,n}(\kappa)&=&\tilde{a}_{n,m}(\kappa) = c_n(\kappa)\,
c_{m+n+\kappa}(\kappa), \\
c_m(\kappa) &=& \frac{e(m)}{e(\kappa+m)},\qquad
e(m)~ =~ \frac{m}{m-1+\kappa}.
\end{eqnarray}
In the next section, we will build
the polynomials with the help of these recursion relations.

%
%Note that $\Delta ^\kappa $ is self-adjoint in the space of 
%functions $f(z,\bar z)$ with the norm \cite{Ko} 
%\[ \Vert f\Vert _\kappa ^2=\int _D\,|f(z,\bar z)|^2\,(w(z,\bar z))^\kappa 
%\,dz\,d\bar z,\,\,\kappa >-\frac{1}{3},\]
%\[ w(z,\bar z)=-z^2\bar z^2+4z^3+4\bar z^3-18z\bar z+27, \]
%where $D$ is a bounded domain defined by the curve $w(z,\bar z)=0$. 
%

\section{Results}
\setcounter{equation}{0}

As a first step towards the complete solution, it is instructive to
compute the simpler $P^\kappa_{m,0}$ polynomials, which were considered 
first by Jack \cite{Jac} (see also \cite{Pr}). Combining the recursion 
relations (\ref{rec-z1}) and (\ref{rec-z2}), we get
\be
P^\kappa_{m+1,0} = z_1 \, P^\kappa_{m,0}-c_m \, z_2 \, 
P^\kappa_{m-1,0}+d_m \, P^\kappa_{m-2,0},
\label{rec-mzero}
\ee
where $d_m=c_m \, c_{m-1} \, c_{m-1+\kappa}$ (for brevity we drop the
$\kappa$ dependence in $c_m$).
From the general structure (\ref{P-struct}) of $P^\kappa_{m,n}$,
it is natural to decompose $P^\kappa_{m,0}$ as
\be
P^\kappa_{m,0} = \sum_{l=0}^{[\frac{m}{3}]} z_1^{m-3l}\, 
Q_l^{\kappa,m} (y),\qquad y=\frac{z_2}{z_1^2},
\ee
 $Q_l^{\kappa,m}(y)$ being a polynomial in $y$. Then the recursion 
relation (\ref{rec-mzero}) implies that these $Q_l^{\kappa,m}$ satisfy 
\begin{eqnarray}
Q_0^{\kappa,m+1}(y) & = & Q_0^{\kappa,m}(y) - c_m \,y\,
Q_0^{\kappa,m-1}(y), \\
Q_l^{\kappa,m+1}(y) & = & Q_l^{\kappa,m}(y) - c_m\, y \,
Q_l^{\kappa,m-1}(y) + d_m\, Q_{l-1}^{\kappa,m-1}(y).
\label{rec-l}
\end{eqnarray}
The first relation involves only $Q_l^{\kappa,m}$ with $l=0$ and can
be readily solved with the help of the Gegenbauer polynomials 
$C^\kappa_m(t)$ as
\begin{eqnarray}
Q_0^{\kappa,m} (y)&=&\sum_{i=0}^{[\frac{m}{2}]}\,\frac{(-1)^i}{i!}\,
\frac{m!}{(m-2i)!}\,\frac{\Gamma(\kappa+m-i)}{\Gamma(\kappa+m)} \, y^i
\nonumber \\
%&=& e(m)_m \, y^{m/2} 
&=& \frac{y^{m/2}}{e(m+1)_{-m}}
\, C^\kappa_m \left( \frac{1}{2 \sqrt{y}}\right) ,
\end{eqnarray}
where $1/e(x+1)_{-i}$ denotes the product\footnote{Similarly, for
positive $i$, $e(x)_i = e(x)\,e(x+1)\cdots e(x+i-1)$.  This is a
functional generalization of the Pochhammer symbol $(x)_i =
\Gamma(x+i)/\Gamma(x), i\in \Z$ used later in the
text.} $e(x)\,e(x-1)\cdots e(x-i+1)$.
For higher $l$, we try the following ansatz
\be
Q_l^{\kappa,m}(y) = \alpha_l^m \, Q_{0}^{\kappa+l,m-3l}(y),
\ee
which solves (\ref{rec-l}), provided that the constants $\alpha_l^m$
are
\be
\alpha_l^m = \frac{m!}{l!\,(m-3l)!}\,\frac{\Gamma(\kappa+m-2l)}
{\Gamma(\kappa+m)}.
\ee
Therefore, we conclude that the polynomials $P^\kappa_{m,0}(z_1,z_2)$
are just some particular linear combinations of the one-variable 
Gegenbauer polynomials.  One obtains the other set of polynomials
$P^\kappa_{0,n}$ using the relation $P^\kappa_{0,n}(z_1,z_2)=
P^\kappa_{n,0}(z_2,z_1)$.

The recursion relation (\ref{rec-mzero}) is also very useful to derive
a generating function for the $P^\kappa_{m,0}$ polynomials.  Indeed,
plugging in (\ref{rec-mzero}) the following function
\be
G^\kappa_0(u) = \sum_{m=0}^\infty
%\frac{u^m}{e(m)_m}
e(m+1)_{-m}\, u^m
\,P^\kappa_{m,0},
\label{gen-mzero}
\ee
we obtain the first-order differential equation,
easily solved by
\be
G^\kappa_0(u) = (1-z_1 u+z_2 u^2-u^3)^{-\kappa}.
\ee

This generating function is perfectly suited to prove some basic
properties of these polynomials, such as
\be
\partial_1 P^\kappa_{m,0} = m\,P^{\kappa+1}_{m-1,0}, \qquad
\partial_2 P^\kappa_{m,0} = -\,\frac{m(m-1)}{\kappa+m-1}\,
P^{\kappa+1}_{m-2,0}.
\ee

We will build the general polynomials with the help of the
$P^\kappa_{m,0}$, using the property
\be
P^\kappa_{m,0}\,P^\kappa_{0,n} = \sum_{i=0}^{{\rm min}(m,n)}
\gamma^i_{m,n}\,
P^\kappa_{m-i,n-i}.
\label{ppgammap}
\ee
This is a consequence of Eq. (\ref{main}), with the notable 
difference that the sum on the right-hand side is over a restricted domain 
(actually, it parallels exactly the $SU(3)$ Clebsch--Gordan
decomposition).

For the proof, we proceed by iteration, assuming that
(\ref{ppgammap}) is valid up to $(m,n)$.  Then, with repeated use of
(\ref{rec-z1}) and (\ref{rec-z2}), we get
\be
P^\kappa_{m,0}\,P^\kappa_{0,n+1} = \sum_{i=0}^{{\rm min}(m,n+1)}
\gamma^i_{m,n+1}\,P^\kappa_{m-i,n+1-i} + c_n\,\delta^i_{m,n+1}\,
P^\kappa_{m+1-i,n-1-i},
\label{ppgammadeltap}
\ee
where we defined
\begin{eqnarray}
\gamma^i_{m,n+1} &=& \gamma^i_{m,n} + \tilde{a}_{m-i+1,n-i+1} \, 
\gamma^{i-1}_{m,n} - c_n \, c_{m-i+1} \, \gamma^{i-1}_{m,n-1},
\label{gammadelta1} \\
\delta^i_{m,n+1} &=& c_n^{-1} \, c_{n-i} \, \gamma^i_{m,n} -
\gamma^i_{m,n-1} - a_{m-i+1,n-i} \, \gamma^{i-1}_{m,n-1} \nonumber \\
& & \mbox{} + c_{n-1} \, c_{\kappa+n-1} \, \gamma^{i-1}_{m,n-2}.
\label{gammadelta2}
\end{eqnarray}
From the polynomials normalisation, we already know that
$\gamma^0_{m,n}=1$, and after a straigthforward computation, the
solution to (\ref{gammadelta1}) is found to be
\be
\gamma^i_{m,n} =
%\frac{e(m)_i \, e(n)_i}{e(i)_i \, e(2\kappa+m+n-i)_i}
\frac{e(2\kappa+m+n+1-i)_{-i}}{e(1)_i \, e(m+1)_{-i}\, e(n+1)_{-i}},
\label{sol-gamma}
\ee
which implies that $\delta^i_{m,n+1}=0$ in (\ref{gammadelta2}).
%The recursion over the other index
%$m$ is proved using the symmetry under exchange of $m,n$ and $z_1,z_2$.

The constructive aspect of this formula lies in its inverted form.

\medskip\noindent
{\bf Theorem 1}. The Jack polynomials $P_{m,n}^\kappa$
of type $A_2$ are given by the formula
\be
P^\kappa_{m,n} = \sum_{i=0}^{{\rm min}(m,n)} \beta^i_{m,n}\,
P^\kappa_{m-i,0}\,P^\kappa_{0,n-i},
\label{ppbetap}
\ee
where the constants are
\be
\beta^i_{m,n} = 
\frac{(-1)^i}{i!\,(\kappa+1)_{-i}} \,\frac{3\kappa+m+n-2i}{3\kappa+m+n-i}
\,
%\frac{(m)_i\,(n)_i\,(\kappa)_i\,(3\kappa+m+n-1)_i} 
%{(\kappa+m-1)_i\,(\kappa+n-1)_i\,(2\kappa+m+n-1)_i}.
\frac{(\kappa+m)_{-i} \,(\kappa+n)_{-i} \,(2\kappa+m+n)_{-i}}
{(m+1)_{-i}\, (n+1)_{-i}\, (3\kappa+m+n)_{-i}}.
\ee
Note that the $\beta^i_{m,n}$ are obtained using the relation
\be 
\beta _{m,n}^i=-\,\sum_{j=0}^{i-1}
\beta^j_{m,n}\,\gamma^{i-j}_{m-j,n-j}.
\ee
From this theorem, we see that the construction of a general
polynomial $P^\kappa_{m,n}$ is similar to the construction of
 $SU(3)$ representations from tensor
products of the two fundamental representations. 

Likewise, one can explicitly study other types of decompositions,
such as
\be
P^\kappa_{m,0}\,P^\kappa_{n,0} = \sum_{i=0}^{{\rm min}(m,n)} \tilde 
\gamma ^i_{m,n}\,P^\kappa_{m+n-2i,i},
\label{ppgammatp}
\ee
with the coefficients
\be
\tilde \gamma ^i_{m,n} = 
%\frac{e(m)_i \, e(n)_i}{e(i)_i \,  e(\kappa+m+n-i)_i}.
\frac{e(\kappa+m+n+1-i)_{-i}}{e(1)_i\, e(m+1)_{-i} \,e(n+1)_{-i}}.
\ee
The proof is essentially the same as for (\ref{ppgammap}).  Here
again, the summation on the right-hand side is on a restricted domain, 
compared to (\ref{main}).

\medskip\noindent
{\bf Theorem 2}. There is another formula for polynomials
$P_{m,n}^\kappa$ at $m\geq n$:
\be
\tilde\gamma^n_{m+n,n}\,P_{m,n}^\kappa =\sum _{i=0}^n\,\tilde 
\beta _{mn}^i\,P^\kappa_{m+n+i,0}\,P^\kappa_{n-i,0},
\ee
where
\be
\tilde \beta _{m,n}^i=
%(-1)^i\,\frac{(\kappa )_i}{i!}
%\frac{(\kappa+m+n)^i}{(m+n+1)^i}
%\frac{(m)^i}{(\kappa+m+1)^i}
%\frac{(n)_i}{(\kappa+n-1)_i}
%\frac{m+2i}{m}.
\frac{(-1)^i}{i!\, (\kappa+1)_{-i}}
\frac{m+2i}{m}
\frac{(\kappa+m+n)_i}{(m+n+1)_i}
\frac{(m)_i}{(\kappa+m+1)_i}
\frac{(\kappa+n)_{-i}}{(n+1)_{-i}}.
\ee

This theorem is a simple consequence of (\ref{ppgammatp}), and the
coefficients $\tilde\beta^i_{m,n}$ are found using
\be
\tilde\beta^i_{m,n} =
-\left(\tilde\gamma^{n-i}_{m+n+i,n-i}\right)^{-1}\,
\sum_{j=0}^{i-1} \, \tilde\beta^j_{m,n} \, \tilde 
\gamma^{n-i}_{m+n+j,n-j} .
\ee

As a by-product of (\ref{ppbetap}), specializing it to the case
$\kappa=1$, where $P^\kappa_{m,n}$ are
nothing but the $SU(3)$ characters, we get
\be
P^{1}_{m,n}= P^{1}_{m,0}\,P^{1}_{0,n}-P^{1}_{m-1,0}\,P^{1}_{0,n-1}. 
\ee
From this we easily deduce the generating function for $SU(3)$
characters (see e.g. \cite{PS})
\be
G^{1}(u,v)=\sum_{m,n=0}^\infty u^m\,v^n\,P^{1}_{m,n} =
\frac{1-uv}{(1-z_1\,u+z_2\,u^2-u^3)\,(1-z_2\,v+z_1\,v^2-v^3)}.
\ee

\section{Conclusion}
\setcounter{equation}{0}

In this letter we have solved the quantum three-body 
Calogero--Sutherland model exactly.  The wave functions are known to
be Jack polynomials, and our construction gives explicit
expansion of them. They appear to be constructed with
Gegenbauer polynomials. 

Since the wave functions correspond, for special values of $\kappa$,
to zonal spherical polynomials, we have obtained, as a
by-product,  explicit expression for zonal spherical 
functions of the symmetric spaces $SU(3)/SO(3)$ ($\kappa =1/2$), 
$SU(3)\times SU(3)/SU(3)$ ($\kappa =1$), $SU(6)/Sp(3)$ ($\kappa =2$), 
and $E_{6(-78)}/F_4$ ($\kappa =4$).

Due to the algebraic framework, many aspects of this work can be
applied to the $N$-body model, for instance Eq. (\ref{ppgammap}) is 
easy to generalize to the $SU(N)$ case.

Let us also remark that preliminary investigations indicate that 
relations similar to
(\ref{rec-z1}) hold in the case of the Macdonald polynomials.

\section*{Appendix: Explicit expressions for $P_{mn}^\kappa$ 
with $m+n\leq 4$}
\setcounter{equation}{0}
In addition to those already given in the main text, we list here the
first few polynomials $P_{m,n}^\kappa$ with $m+n \leq 4$:
%\be
\begin{eqnarray}%{lcl}
P_{2,0}^\kappa &=& z_1^2 - \frac{2}{\kappa+1} z_2 \nonumber \\
P_{1,1}^\kappa &=& z_1 z_2 -\frac{3}{2\kappa+1} \nonumber \\
P_{3,0}^\kappa &=& z_1^3 - \frac{6}{\kappa+2} z_1 z_2 + \frac{6}
{(\kappa+1)(\kappa+2)}\nonumber \\
P_{2,1}^\kappa &=& z_1^2 z_2 -\frac{2}{\kappa+1} z_2^2 -
\frac{3\kappa+1}{(\kappa+1)^2} z_1 \nonumber \\
P_{4,0}^\kappa &=& z_1^4 - \frac{12}{\kappa+3} z_1^2 z_2 + \frac{12}
{(\kappa+2)(\kappa+3)} z_2^2 + \frac{24}{(\kappa+2)(\kappa+3)} z_1
\nonumber \\
P_{3,1}^\kappa &=& z_1^3 z_2 - \frac{6}{\kappa+2} z_1 z_2^2 -
\frac{3(3\kappa+2)}{(\kappa+2)(2\kappa+3)} z_1^2
+\frac{30}{(\kappa+2)(2\kappa+3)} z_2 \nonumber \\
P_{2,2}^\kappa &=& z_1^2 z_2^2 - \frac{2}{\kappa+1} (z_1^3 + z_2^3)
 - \frac{12(\kappa-1)}{(\kappa+1)(2\kappa+3)}
z_1 z_2 + \frac{9(\kappa-1)}{(\kappa+1)^2(2\kappa+3)}\nonumber
\end{eqnarray}
%\ee

\indent

\section*{Acknowledgements} One of the authors (A.P.) would like to thank 
Prof.~P.~Sorba and 
the Laboratoire de Physique Th{\'e}orique LAPTH for hospitality.

%%%%%%

\end{document}